# Frequency-induced Negative Magnetic Susceptibility in Epoxy/Magnetite Nanocomposites

**Che-Hao Chang[1], Shih-Chieh Su[2], Tsun-Hsu Chang[1, 2*], & Ching-Ray Chang[3**]**

[1]Interdisciplinary Program of Sciences, National Tsing Hua University, Hsinchu, Taiwan
[2]Department of Physics, National Tsing Hua University, Hsinchu, Taiwan
[3]Department of Physics, National Taiwan University, Taipei, Taiwan

*To whom the correspondence should be address: thschang@phys.nthu.edu.tw
**To whom the correspondence should be address: crchang@phys.ntu.edu.tw

## ABSTRACT

The epoxy/magnetite nanocomposites express superparamagnetism under a static or a low-frequency electromagnetic field. At the microwave frequency, said the X-band, the nanocomposites reveal an unexpected diamagnetism. To explain the intriguing phenomenon, we revisit the Debye relaxation law with the memory effect. The magnetization vector of the magnetite is unable to synchronize with the rapidly changing magnetic field, and it contributes to superdiamagnetism, a negative magnetic susceptibility for nanoparticles. The model just developed and the fitting result can not only be used to explain the experimental data in the X-band, but also can be used to estimate the transition frequency between superparamagnetism and superdiamagnetism.

## Introduction

The paramagnetic materials will be aligned with the direction of the externally applied magnetic field, and hence the real part of the susceptibility will be positive. The diamagnetic materials will be arrayed in the opposite direction, and hence result in the negative real susceptibility. Most metals, like copper and silver, will express a weak diamagnetism since the averaged orbital angular momentum of electrons will be in the opposite direction of the magnetic field, called the Langevin diamagnetism[1-2]. The generated magnetic susceptibility is usually very small, around $10^{-4}$-$10^{-5}$. For magnetic materials with small particle sizes, the superparamagnetism will be observed from the thermal agitation[3-6]. The epoxy/magnetite ($Fe_3O_4$) nanocomposites show superparamagnetism as expected at low frequencies[3-6], but our experiment revealed that the $Fe_3O_4$ nano-powder also exhibits peculiar diamagnetism in the X-band. A similar phenomenon was observed on $La_{0.7}Sr_{0.3}MnO_3$ nonocomposites[7] and $FeNi_3$/C nanocapsules[8]. As the order of the negative susceptibility of those materials is much higher than $10^{-4}$ from the Langevin diamagnetism, there must be a different mechanism accounting for the transition from paramagnetism to diamagnetism in our epoxy/magnetite ($Fe_3O_4$) nanocomposites.

The electromagnetic properties of nano-materials are different from their bulk counterpart. A well-known example is that the color of gold particles depends on the sizes[9]. Besides, the ferromagnetic particles, like magnetite, display superparamagnetism when the size is in the nanoscale. The superparamagnetic effect is similar to the paramagnetic effect but with much higher magnetic susceptibilities[10]. The existence of superparamagnetic property arises since the relaxation time scale of the nanoparticles becomes much smaller than usual due to the reduction of the particle volume[11-13]. However, this work shows that the superparamagnetic property become diamagnetism with high magnetic susceptibilities which is at least 4 orders of magnitude higher than the Langevin diamagnetism at high frequencies as the reason we call this phenomenon superdiamagnetism. Under usual circumstances, the permeability of superparamagnetic particles can be explained by the Debye relaxation model[14], such as the FeAl@(Al, Fe)$_2$O$_3$ nanoparticles[15], the Fe/Ag/citrate nano-composites in 2-18 GHz[16], and magnetite in kHz frequency range[17]. Nevertheless, the Debye relaxation model fails to explain the negative susceptibility, and hence an amended physics mechanism is needed.

Depending on the structure of the materials, the spins will naturally be in stable states, that is, in the directions that make the free energy reach a minimum. For a nanoscale material, the thermal fluctuation allows the spins to transit between stable states with the volume reduction of energy barriers. The transition of states contributes to the magnetic susceptibility of superparamagnetic particles. The work of Klik *et al.*[18] shows that the transition of states may contribute to the negative magnetic susceptibility by considering the memory effect of nanoparticles. They considered the correction of the master equation of the spins by an exponential memory kernel. Reference [18] predicts that the material with the uniaxial anisotropy will express diamagnetism when the frequency is greater than a threshold frequency. However, there is no detailed analysis of the physics origins of the high-frequency diamagnetism and also no experimental evidence to support the prediction of the memory induced diamagnetism.

Moreover, Ref. [18] focused their discussion on uniaxial particles having just two stable states. The material of interest (i.e., the magnetite) carries the cubic anisotropy which has six or eight states, depending on the sign of anisotropy constant[19]. As a result, the nano $Fe_3O_4$ has three distinct transition rates with multiple relations[20]. Nevertheless, when it comes to the permeability, the lowest transition rate dominates the magnetization. This work proposes a model considering the cubic anisotropy, and the results agree well with our experimental findings.

## Results

### Permittivity

The transmission/reflection methods can characterize materials' electromagnetic (EM) properties over a broad frequency range. The measured transmission/reflection coefficients using a network analyzer uniquely determine the complex permittivity $\varepsilon$ and the complex permeability $\mu$, when the sample thickness $d$ is smaller than a quarter of the guide wavelength ($\lambda_\mathrm{g}$)[21]. Here, we adapt the relative permittivity $\varepsilon/\varepsilon_0\,(=\varepsilon'+i\varepsilon'')$ and permeability $\mu/\mu_0\,(=\mu'+i\mu'')$. $\varepsilon_0$ and $\mu_0$ denote the permittivity and the permeability of vacuum. Four different volume fractions of epoxy/magnetite composites (0%, 6%, 12%, and 18%) are measured. Then, the effective medium theory is introduced to extract the EM properties of the nano $Fe_3O_4$ powder.

While there are several different effective medium theories, Chang *et al.* discussed three different models and concluded that the Looyenga model[22-23] is a suitable model to fit the permittivity of the epoxy/$Fe_3O_4$ nanocomposites[3]. It reads,

$$\varepsilon_\mathrm{eff}^{1/3} = (1-v_f)\varepsilon_\mathrm{h}^{1/3} + v_\mathrm{f}\,\varepsilon_\mathrm{f}^{1/3}, \tag{1}$$

where $\varepsilon_\mathrm{eff}$, $\varepsilon_\mathrm{h}$ and $\varepsilon_\mathrm{f}$ are the permittivity of the composites, the host medium (epoxy), and the filled material ($Fe_3O_4$), respectively. $v_\mathrm{f}$ is the volume fraction of the filled materials.

The measured effective complex permittivity of nanocomposites with different volume fractions are presented in Fig. 1(a). In our experiment, epoxy was chosen to be the host medium. It carries relative permittivity $3.1+0.1i$ and relative permeability near 1.0 with high stability within a broad frequency range. The EM properties of epoxy are confirmed by our transmission/reflection method. We calculate the corresponding $\varepsilon_\mathrm{f}$ based on Eq. (1). The extracted $\varepsilon_\mathrm{f}$ is shown in Fig. 1(b). The extracted complex permittivities $\varepsilon_\mathrm{f}$ differ slightly for different volume fractions $v_f$. The difference might be attributed to the error of the measured volume fraction $v_\mathrm{f}$.

## Permeability

The measured permeability of nanocomposites with different volume fractions are presented in Fig. 2(a). As mentioned, the permeability of pure epoxy is near 1.0, so the magnetic susceptibility comes strictly from the existence of $Fe_3O_4$. To explain the data, we reasonably assume that the magnetic susceptibility is linearly proportional to the volume fraction of the nano $Fe_3O_4$ powder. That is,

$$\chi'_{\mathrm{eff}} + i\chi''_{\mathrm{eff}} = v_{\mathrm{f}}(\chi'_{\mathrm{f}} + i\chi''_{\mathrm{f}}), \tag{2}$$

where $\chi'_{\mathrm{eff}} + i\chi''_{\mathrm{eff}}$ and $\chi'_{\mathrm{f}} + i\chi''_{\mathrm{f}}$ are the magnetic susceptibility of the composites and the filled material ($Fe_3O_4$), respectively. The extracted $\mu_{\mathrm{f}}$ using Eq. (2) is shown in Fig. 2(b) with $\mu'_{\mathrm{f}} + i\mu''_{\mathrm{f}} = (1+\chi'_{\mathrm{f}}) + i\chi''_{\mathrm{f}}$.

For the $Fe_3O_4$ structure, the free energy of the particles will be[23]

$$E = KV(\alpha^2\beta^2 + \beta^2\gamma^2 + \gamma^2\alpha^2) - \mu_0 H M_s V\cos\varphi, \tag{3}$$

where $K$ is the anisotropy constant, $V$ is the volume, $\alpha, \beta, \gamma$ are the direction cosines of the magnetization vector along the *x*, *y*, and *z* axes, $M_s$ is the saturated magnetization, $H$ is the amplitude of the applied field, and $\cos\varphi$ is the direction cosine of the magnetization vector along the *H*-field.

The magnetization vector of $Fe_3O_4$ has eight stable states, all having the form $|\alpha| = |\beta| = |\gamma|$. We, therefore, denote the stable states as the form [111] or $[\bar{1}\bar{1}\bar{1}]$. The numbers are used to represent the ratios of direction cosines, and we put a bar above the first, second, or third number if $\alpha, \beta$ or $\gamma$ is negative. To describe the susceptibility, we define $n_1, n_2, n_3, n_4, n_5, n_6, n_7$, and $n_8$ as the probability of occupation of [111], $[\bar{1}11]$, $[1\bar{1}1]$, $[11\bar{1}]$, $[\bar{1}\bar{1}\bar{1}]$, $[1\bar{1}\bar{1}]$, $[\bar{1}1\bar{1}]$, and $[\bar{1}\bar{1}1]$. Since the antiparallel occupation will cancel each other out, we can further simplify the expression by defining $m_1 = n_1 - n_5$, $m_2 = n_2 - n_6$, $m_3 = n_3 - n_7$, and $m_4 = n_4 - n_8$. The four variables give us enough information to get the magnetization.

We consider the case that direct transitions happen only between two adjacent states[16]. When considering the memory effect, the master equation of $\mathbf{m} = [m_1 \quad m_2 \quad m_3 \quad m_4]^T$ is

$$\Theta\ddot{\mathbf{m}} + \dot{\mathbf{m}} = \vec{\vec{\mathbf{f}}}_{\mathbf{0}}\mathbf{m} + qh\Gamma\mathbf{v} \text{ with } \vec{\vec{\mathbf{f}}}_{\mathbf{0}} = \Gamma\begin{bmatrix} -3 & 1 & 1 & 1 \\ 1 & -3 & -1 & -1 \\ 1 & -1 & -3 & -1 \\ 1 & -1 & -1 & -3 \end{bmatrix}, \tag{4}$$

where $\Theta$ is the memory time and $\Gamma = \Gamma_0 \exp(-q/12)$. A detailed derivation of how we deduce Eq. (4) from the exponential memory kernel can be found in the Supplementary information Part I. $\Gamma_0$ is a constant in the unit of frequency[24], representing the transition frequency under the high-temperature limit and $q = |K|V/(k_B T)$. The term $\exp(-q/12)$ comes from the energy barrier, like the Arrhenius equation. In the ground states, the particle carries free energy $KV/3$. When the magnetization vector transits from a stable state to the others, it will go through a saddle point as the form of $[110]$ with energy $KV/4$, and therefore, the energy barrier is $-KV/12$. For the second term at the right-hand side, $h = \mu_0 M_s H/(-K)$ and $\mathbf{v}$ can be expressed as

$$\mathbf{v}=\frac{1}{4}\begin{bmatrix}3 & -1 & -1 & -1\\ -1 & 3 & 1 & 1\\ -1 & 1 & 3 & 1\\ -1 & 1 & 1 & 3\end{bmatrix}\begin{bmatrix}\cos\varphi_1\\ \cos\varphi_2\\ \cos\varphi_3\\ \cos\varphi_4\end{bmatrix}, \tag{5}$$

where $\cos\varphi_1, \cos\varphi_2, \cos\varphi_3,$ and $\cos\varphi_4$ are the direction cosines of the magnetization vector along [111], $\left[\bar{1}11\right], \left[1\bar{1}1\right]$ and $\left[11\bar{1}\right]$. When there is an externally applied *H*-field, the transition coefficients will change, since the energy of stable states change from $KV/3$ to $KV/3+\mu_0 HM_s V\cos\varphi$ and therefore the exponential terms of the transition rates change. Since the externally applied *H*-field is small in our experiment, we only need the influences of the first order of $h$. The second term at the right-hand side in Eq. (4) comes from the change of the transition rates.

If $h=h_0\exp(-i\omega t)$ with $h_0$ as the amplitude and $\omega$ as the oscillating frequency, we can expect that $\mathbf{m}$ will oscillate with the same frequency. Therefore, we can do the Fourier transformation for the left side. The $\ddot{\mathbf{f}}_{\mathbf{0}}$ has an eigenvalue $-6\Gamma$ corresponding to the eigenvector $[1 \quad -1 \quad -1 \quad -1]^T$. In addition, the $\ddot{\mathbf{f}}_{\mathbf{0}}$ has another eigenvalue $-2\Gamma$ associated with three degenerate eigenvectors $\left[1 \quad 1 \quad 0 \quad 0\right]^T$, $\left[1 \quad 0 \quad 1 \quad 0\right]^T$, and $\left[1 \quad 0 \quad 0 \quad 1\right]^T$. The physical interpretation of eigenvalues and eigenvectors can be found in the Supplementary information Part II. After the matrix operation, Eq. (4) becomes

$$\begin{bmatrix}(6\Gamma-\Theta\omega^2-i\omega)(m_1-m_2-m_3-m_4)\\ (2\Gamma-\Theta\omega^2-i\omega)(m_1+3m_2-m_3-m_4)\\ (2\Gamma-\Theta\omega^2-i\omega)(m_1-m_2+3m_3-m_4)\\ (2\Gamma-\Theta\omega^2-i\omega)(m_1-m_2-m_3+3m_4)\end{bmatrix}=qh\Gamma\begin{bmatrix}1 & -1 & -1 & -1\\ 1 & 3 & -1 & -1\\ 1 & -1 & 3 & -1\\ 1 & -1 & -1 & 3\end{bmatrix}\mathbf{v}. \tag{6}$$

To get magnetization $M$, we define the reduced magnetization $m_r=M/M_s$, and then the value of $m_r$ is related to $m_i$'s by

$$m_r=[\cos\varphi_1 \quad \cos\varphi_2 \quad \cos\varphi_3 \quad \cos\varphi_4]\begin{bmatrix}m_1\\ m_2\\ m_3\\ m_4\end{bmatrix}. \tag{7}$$

For simplicity, we define $1/\lambda_1=6\Gamma-\Theta\omega^2-i\omega$ and $1/\lambda_2=2\Gamma-\Theta\omega^2-i\omega$. Then, we get

$$m_r=[\cos\varphi_1 \quad \cos\varphi_2 \quad \cos\varphi_3 \quad \cos\varphi_4]\begin{bmatrix}1 & -1 & -1 & -1\\ 1 & 3 & -1 & -1\\ 1 & -1 & 3 & -1\\ 1 & -1 & -1 & 3\end{bmatrix}^{-1}\begin{bmatrix}\lambda_1 & 0 & 0 & 0\\ 0 & \lambda_2 & 0 & 0\\ 0 & 0 & \lambda_2 & 0\\ 0 & 0 & 0 & \lambda_2\end{bmatrix}\begin{bmatrix}1 & -1 & -1 & -1\\ 1 & 3 & -1 & -1\\ 1 & -1 & 3 & -1\\ 1 & -1 & -1 & 3\end{bmatrix}qh\Gamma\mathbf{v}. \tag{8}$$

Equations (6) and (8) give us $m_r$ for any direction of the externally magnetic field. However, the orientation is randomly distributed in the experiment. Therefore what we need is the averaged $m_r$ for all the possible directions, denoted as $\langle m_r\rangle$. All $\langle\cos\varphi_n m_n\rangle$ are equal by symmetry, where $n$ = 1, 2, 3, or 4. Therefore we only need to calculate $4\langle\cos\varphi_1 m_1\rangle$.

$$\langle m_r\rangle=[1 \quad 0 \quad 0 \quad 0]\begin{bmatrix}1 & 1 & 1 & 1\\ -1 & 1 & 0 & 0\\ -1 & 0 & 1 & 0\\ -1 & 0 & 0 & 1\end{bmatrix}\begin{bmatrix}\lambda_1 & 0 & 0 & 0\\ 0 & \lambda_2 & 0 & 0\\ 0 & 0 & \lambda_2 & 0\\ 0 & 0 & 0 & \lambda_2\end{bmatrix}\begin{bmatrix}6 & -6 & -6 & -6\\ 2 & 6 & -2 & -2\\ 2 & -2 & 6 & -2\\ 2 & -2 & -2 & 6\end{bmatrix}\begin{bmatrix}\langle\cos\varphi_1\cos\varphi_1\rangle\\ \langle\cos\varphi_1\cos\varphi_2\rangle\\ \langle\cos\varphi_1\cos\varphi_3\rangle\\ \langle\cos\varphi_1\cos\varphi_4\rangle\end{bmatrix}\frac{qh\Gamma}{4}. \tag{9}$$

To find $\langle \cos\varphi_1 \cos\varphi_n \rangle$ for all $n$'s, we can express the direction of the magnetization vector as $(\sin\theta\sin\phi,\ \sin\theta\cos\phi,\ \cos\theta)$ by setting $(\theta,\ \phi)$ as a coordinate of a unit sphere such that $[111]$ identities to $(1,1,1)/\sqrt{3}$. We can find $\cos\varphi_n$ by calculating the inner product between the magnetization vector and the direction of stable states. The average of $\cos\varphi_1 \cos\varphi_n$ for all possible $(\theta,\ \phi)$ is what we need. The values are $\langle \cos\varphi_1 \cos\varphi_1 \rangle = 2/3$ and $\langle \cos\varphi_1 \cos\varphi_n \rangle = 2/9$ for the rest.

$$\langle m_r \rangle = [\lambda_1 \quad \lambda_2 \quad \lambda_2 \quad \lambda_2] \begin{bmatrix} 6 & -6 & -6 & -6 \\ 2 & 6 & -2 & -2 \\ 2 & -2 & 6 & -2 \\ 2 & -2 & -2 & 6 \end{bmatrix} \begin{bmatrix} 6 \\ 2 \\ 2 \\ 2 \end{bmatrix} \frac{qh\Gamma}{36} = \frac{4}{3}\lambda_2 qh\Gamma. \tag{10}$$

Eventually, we can write the magnetization as

$$M = M_s \langle m_r \rangle = \frac{2}{3} \frac{2\Gamma}{2\Gamma - \Theta\omega^2 - i\omega} \frac{-KV}{k_b T} \frac{\mu_0 M_s^2}{-K} H. \tag{11}$$

Our calculation shows that the only decay rate contributes to the magnetization is $2\Gamma$, and therefore the mathematical form looks just like that for uniaxial particles. This result is not limited to the randomly oriented case as shown in the Supplementary information Part II.

The magnetic susceptibility of $Fe_3O_4$ will be

$$\chi'_f + i\chi''_f = \frac{M}{H} = \frac{\mu_0 M_s^2}{-K} \frac{m}{h} = \frac{2\mu_0 M_s^2 q}{3|K|} \frac{\Gamma_f}{\Gamma_f - \Theta_f f^2 - if}, \tag{12}$$

where we use the frequency $f$ to replace $\omega$, $\Gamma_f = 4\pi\Gamma$, and $\Theta_f = \Theta/2\pi$.

The three unknown variables are $2\mu_0 M_s^2 q/(3|K|)$, $\Gamma_f$, and $\Theta_f$. We have two curves: the real permeability and the imaginary permeability, which give us two conditions. Since we have two conditions but three variables, it will be beneficial if we can obtain another constraint. From Eqs. (2) and (12), the real part and imaginary part of magnetic susceptibility of composites will be

$$\begin{cases} \chi'_{\text{eff}} = \dfrac{2\mu_0 M_s^2 V}{3k_B T} f_v \dfrac{\Gamma_f(\Gamma_f - \Theta_f f^2)}{f^2 + (\Gamma_f - \Theta_f f^2)^2} \\ \chi''_{\text{eff}} = \dfrac{2\mu_0 M_s^2 V}{3k_B T} f_v \dfrac{f\Gamma_f}{f^2 + (\Gamma_f - \Theta_f f^2)^2} \end{cases}. \tag{13}$$

The two equations merge to

$$f \frac{\chi'_{\text{eff}}}{\chi''_{\text{eff}}} = \Gamma_f - \Theta_f f^2. \tag{14}$$

Since the left-hand side can be directly computed from the experimental data, we can get $\Theta_f$ by the slope of the left-hand side term versus $f^2$, as shown in Fig. 3(a). The $\Theta_f$ should be a positive quantity, and we indeed found a decreased line in Fig. 3(a). After getting the $\Theta_f$, one can get $\Gamma_f$ by putting $\Theta_f f^2$ to the left side of the Eq. (14) as shown in Fig. 3(b). Notably, if one checks Fig. 2(a) carefully, the real and imaginary susceptibilities ($\chi'_{\text{eff}}$ and $\chi''_{\text{eff}}$) above 10 GHz are very close to zero. Since $\chi''_{\text{eff}}$ is in the dominator of the left-hand side of Eq. (14), the very small value of $\chi''_{\text{eff}}$ will enlarge the uncertainty. Therefore, the regions to the left of the dashed lines in Figs. 3(a) and 3(b) are more reliable and those are the region of interest. The consistency between Eq. (14) and the measured data suggests the memory model with cubic anisotropic materials works well. After finding $\Theta_f$ and $\Gamma_f$, we can then determine the remaining coefficient in Eq. (12), i.e., $2\mu_0 M_s^2 q/(-3K)$. The fitting results of the three variables are listed in Table 1.

Figure 4(a) shows the magnetic susceptibility of nano-magnetite based on Eq. (12) and Table 1. Using Eq. (12), we can estimate the permeability outside the range of the X-band. As an example, from the form of $\chi_f'$, one can expect that the transition between superparamagnetism and superdiamagnetism happens at $f_t = \sqrt{\Gamma_f/\Theta_f}$. When the frequency is higher than $f_t$, the real part of the magnetic susceptibility becomes negative. This result is similar to the frequency response of an RLC circuit as Eq. (4) can be analogous to an RLC circuit, which is explained in the Supplementary information Part III. $f_t$ will be 5.33 GHz using the parameters in Table 1. The transition frequency $f_t$ agrees with the experimental observation that the epoxy/magnetite nanocomposite expresses superparamagnetism at 2.45 GHz[3] and exhibits superdiamagnetism in the X-band (8-12 GHz, i.e., this work).

## Discussion

We have proposed a memory model with thermal agitation of the ferromagnetic nanoparticle. The superparamagnetism is from the thermal fluctuations between stable states, while the superdiamagnetism originates from the out of phase response with the external driven field (Supplement Part III). The experimental results of the epoxy/magnetite nanocomposites showed the memory effects clearly at microwave frequency. An unexpected negative susceptibility can be at least four orders of magnitude higher than the Langevin diamagnetism (Fig. 4). We just show that the memory effect yields the correct frequency response of the susceptibility for cubic anisotropic materials where the magnetic susceptibility changes from positive at low frequency to negative at high frequency. To demonstrate the importance of the value of $\Theta$, we examine the minimum of the real susceptibility in Eq. (12):

$$\chi'_{\min} = -\frac{2\mu_0 M_s^2 q}{3|K|}\left(\frac{2\Gamma\Theta}{2\sqrt{2\Gamma\Theta}+1}\right), \tag{15}$$

when $\omega_{\min} = \sqrt{(2\Gamma/\Theta) + \sqrt{2\Gamma/\Theta}/\Theta}$. Note that the value of $2\Gamma\Theta$, i.e., the ratio of the memory time ($\Theta$) and the relaxation time $1/2\Gamma$, affects the amplitude of the minimum. Under the limit $2\Gamma\Theta \to 0$, $\chi'_{\min} \to 0$ with the corresponding frequency $\omega \to \infty$, which implies that the memory effect is feeble and the susceptibility will comply with the Debye relaxation formula.

Another extreme case is $2\Gamma\Theta \to \infty$. In this case, $\chi'_{\min}$ is negative associated with extremely large $\chi''$. However, the corresponding frequency $\omega \to 0$ and once the order of frequency is larger than that of the corresponding frequency $\omega_{\min}$, $\chi'$ will be quite close to 0. The reason we can observe the superdiamagnetism phenomenon in X-band is that the memory time scale and relaxation time scale are comparable, and hence superdiamagnetism becomes observable as $\omega \approx \sqrt{\Gamma/\Theta}$.

The diamagnetism expels the magnetic fields within a material. Superconductor is a perfect diamagnetic material with $\chi' = -1$, which results in no internal magnetic field. This study explored the memory effect, which produced a strong diamagnetism for nanocomposites at room temperature. Such a new mechanism is called superdiamagnetism. The superdiamagnetism significantly reduces the ac magnetic field without affecting the ac electric field. It is fundamentally different from the Meissner effect. The limit of the negative susceptibility is yet to be uncovered. Besides, the adjustable magneto-dielectric properties of the composite materials can be used for multilayer antireflection coating or the stealth aircraft/warship. It deserves further theoretical and experimental studies.

## Methods

### Transmission/reflection method

To conduct the transmission/reflection measurement, we sandwich the sample between two adaptors connecting to a performance network analyzer (PNA). The experiment is performed using the standard WR90 waveguide with inner dimensions of 0.9 in by 0.4 in. The two adaptors serve as the two ports, which convert the WR90 waveguide to the 2.4 SMA (SubMiniature version A) coaxial cables. The two ports are calibrated before measurement. The sample is filled in a uniform and hollowed WR90 waveguide. The measured transmission/reflection coefficients are then used to extract the complex permittivity/permeability of samples.

### Preparation of the nanocomposite samples

Figure 5 illustrates the sample preparing procedure:

1. Cleaning the waveguide

The standard WR90 waveguide made of copper should be oxidation-free and clean inside with low surface roughness. The purpose of cleaning the waveguide is to avoid the conductor loss and to ensure that the loss comes strictly from the nanocomposite.

We first soaked the waveguide into the copper polishing solution to remove the oxide. Then, the waveguides were immersed in the distilled water to remove the remnant acid. After that, we bathed the waveguide into the acetone to remove the organic dust and the water. Finally, we soaked it into isopropyl alcohol (IPA) to remove acetone buffer. An ultrasonic cleaning machine is used during the process. Eventually, we obtain a clean waveguide, as shown in Fig. 5.

2. Preparation of filler:

Epoxy resin (epoxy A) mixing with hardener (epoxy B) will become very sticky and soon form a thermosetting polymer. To achieve the nanocomposite with a uniform distribution, we first poured epoxy A into a small jar, and then placed the nano $Fe_3O_4$ powder into it. Then, we stirred the liquid in the jar to make them mix. We would heat the jar to make $Fe_3O_4$ dissolve more easily. Finally, we put the epoxy B into it and kept stirring until the mixture looks evenly distributed.

3. Filling the waveguide

After cleaning the waveguide and preparing the filler, we needed to fill the waveguide with the filler. During the solidifying process, the volume of the epoxy will shrink and may result in an air gap between the sample and the waveguide. To solve this problem, we piled the waveguide and Teflon and then filled the whole pile. After filling the pile, we heated it so that epoxy can solidify. When it solidified several hours later, we removed Teflon, and one would find that the thickness of the solidified filler exceeds the waveguide length. We used a grinding machine to remove the remnants of filler and polish the surface of the waveguide.

After completing the sample preparation procedures, we then went through a scanning electron microscope (SEM) measurement to check the uniformity of the nanocomposite. The results are shown in Fig. 6, which assures the uniformity of samples. The SEM image of the sample is shown in Fig. 6(a). The $Fe_3O_4$ particles are rich in Fe. Using the Fe Ka X-ray microanalysis, we can analyze the distribution of iron, as in Fig. 6(b). Figure 6(c) shows the ingredient analysis using the X-ray emission spectrum.

## Author contributions statement

T. H. conceived the experiment, C.H. and S.C. experimented, C.H., C.R., and T.H. provided theoretical model, C.H., C.R., and T.H. analyzed the results. All authors reviewed the manuscript.

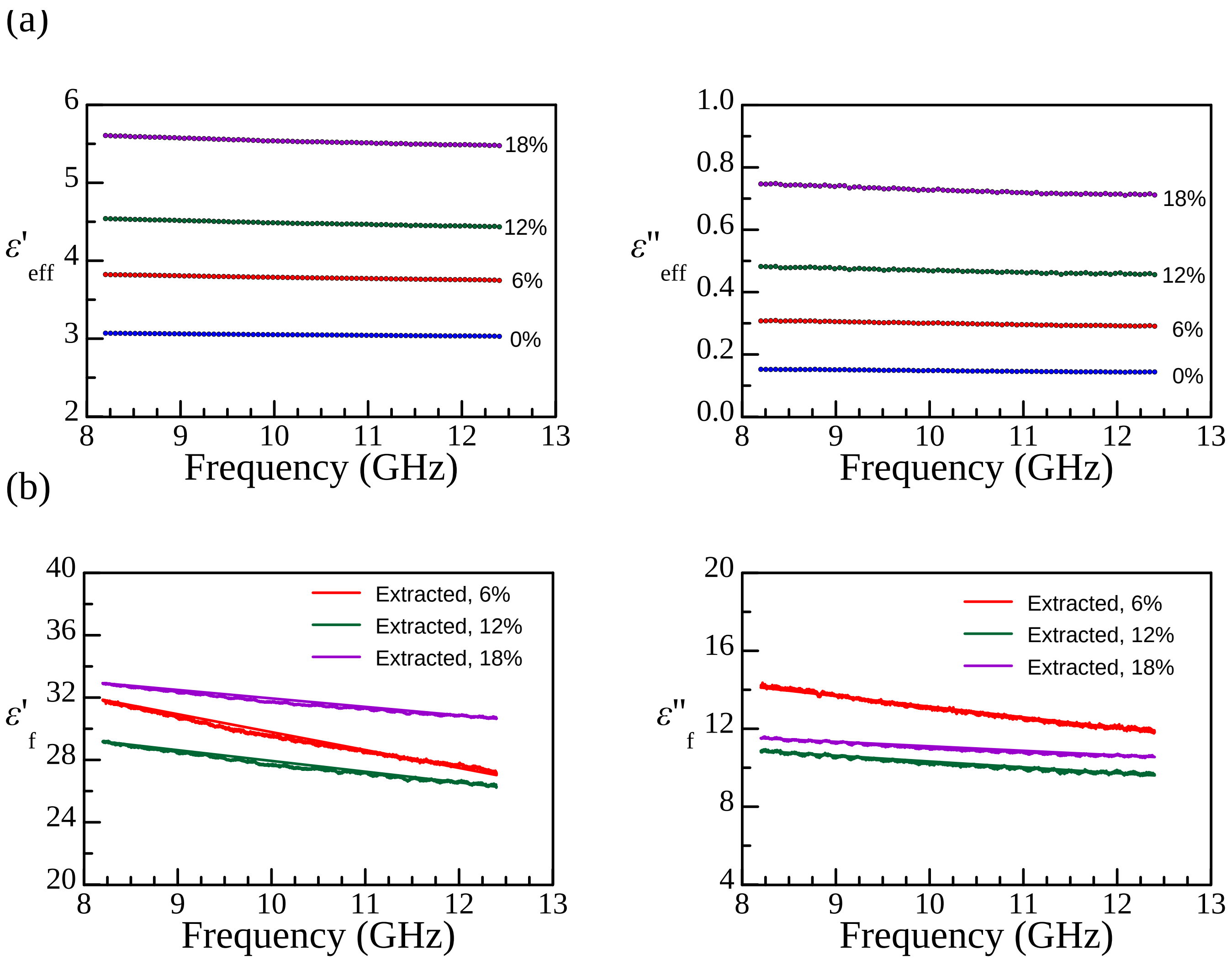

**Figure 1.** (a) The measured (effective) complex permittivity versus frequency and (b) the extracted complex permittivity of the $Fe_3O_4$ nanoparticles using the Looyenga model (Eq. (1)).

(a)

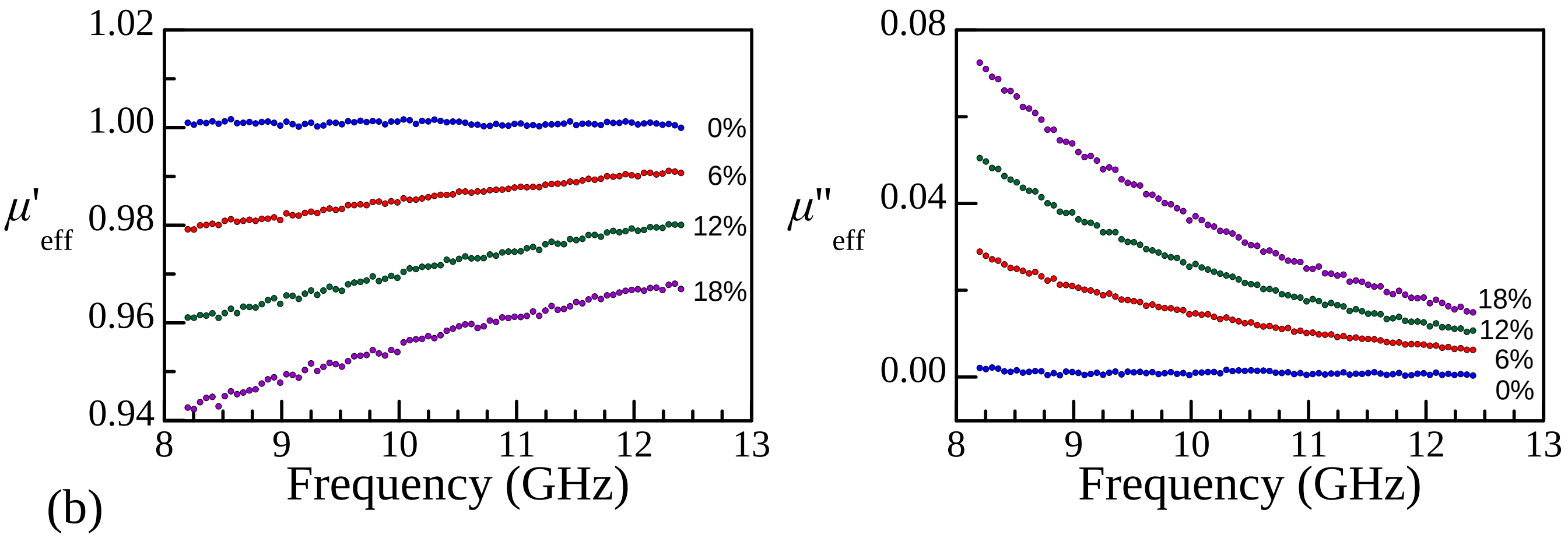

(b)

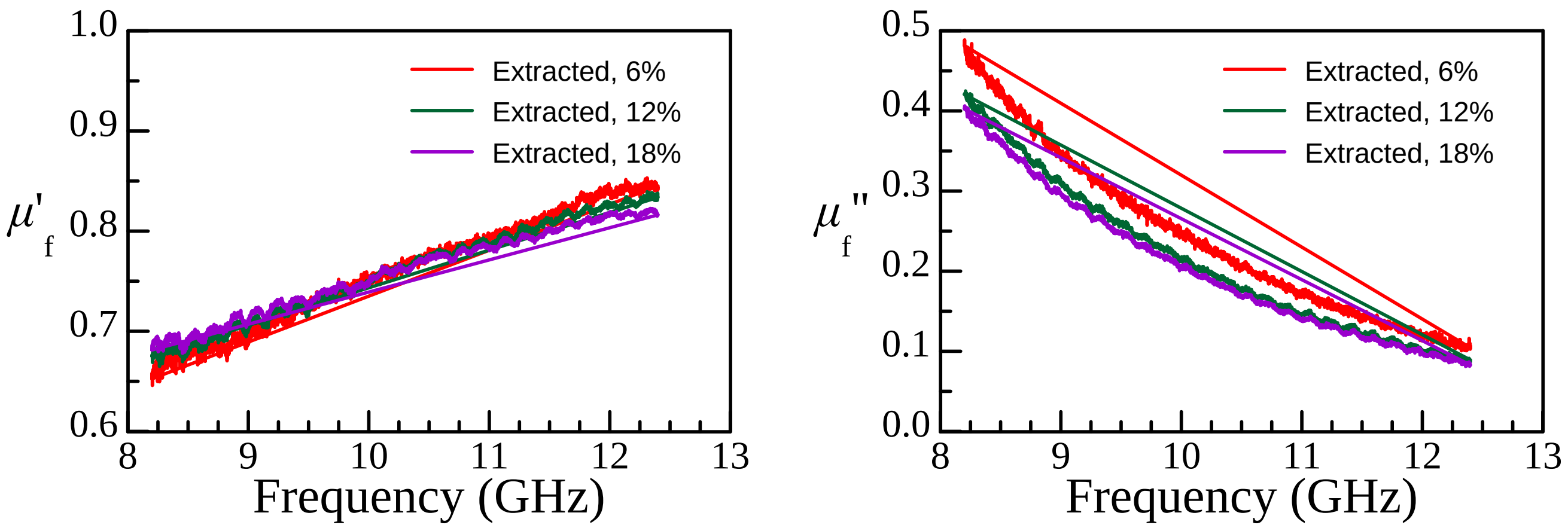

**Figure 2.** (a) The measured (effective) complex permeability versus frequency and (b) the extracted complex permeability of the $Fe_3O_4$ nanoparticles using the linearly proportional model (Eq. (2)).

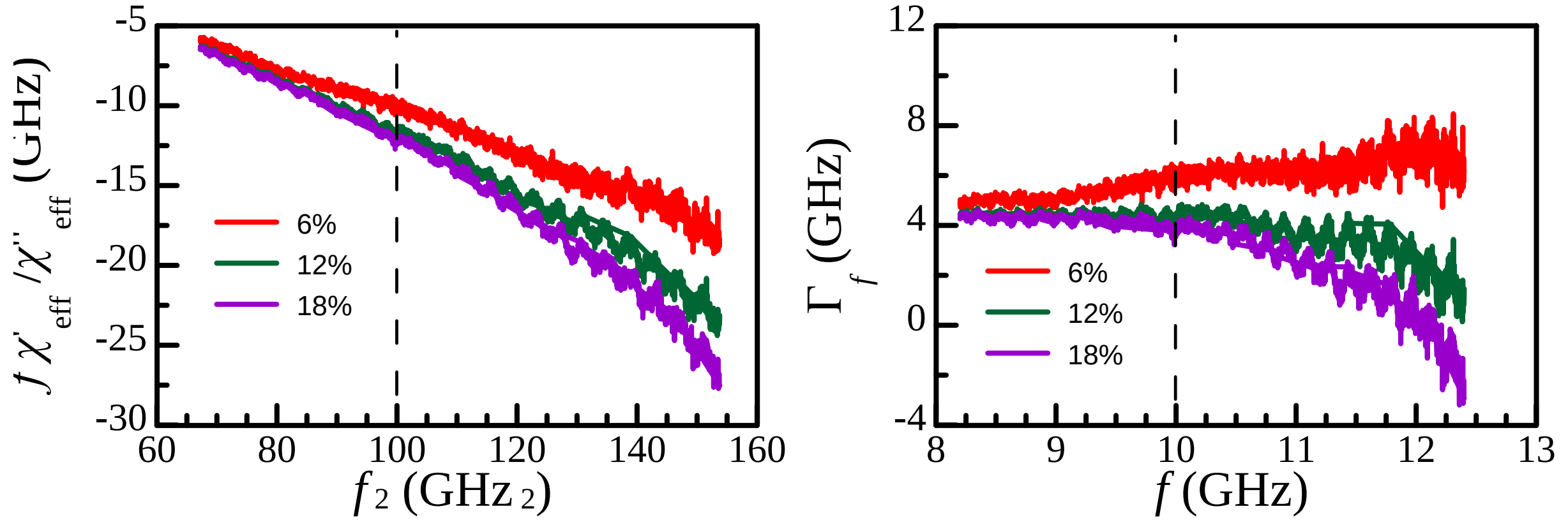

**Figure 3.** (a) From Eq. (14), $\Theta_f$ is associated with the slope of the lines, i.e., $f\chi'_{\text{eff}}/\chi''_{\text{eff}}$ vs. $f^2$. (b) Rewriting Eq. (14), $\Gamma_f = f\chi'_{\text{eff}}/\chi''_{\text{eff}} + \Theta_f f^2$. The value of $\Gamma_f$ is expected to be constant.

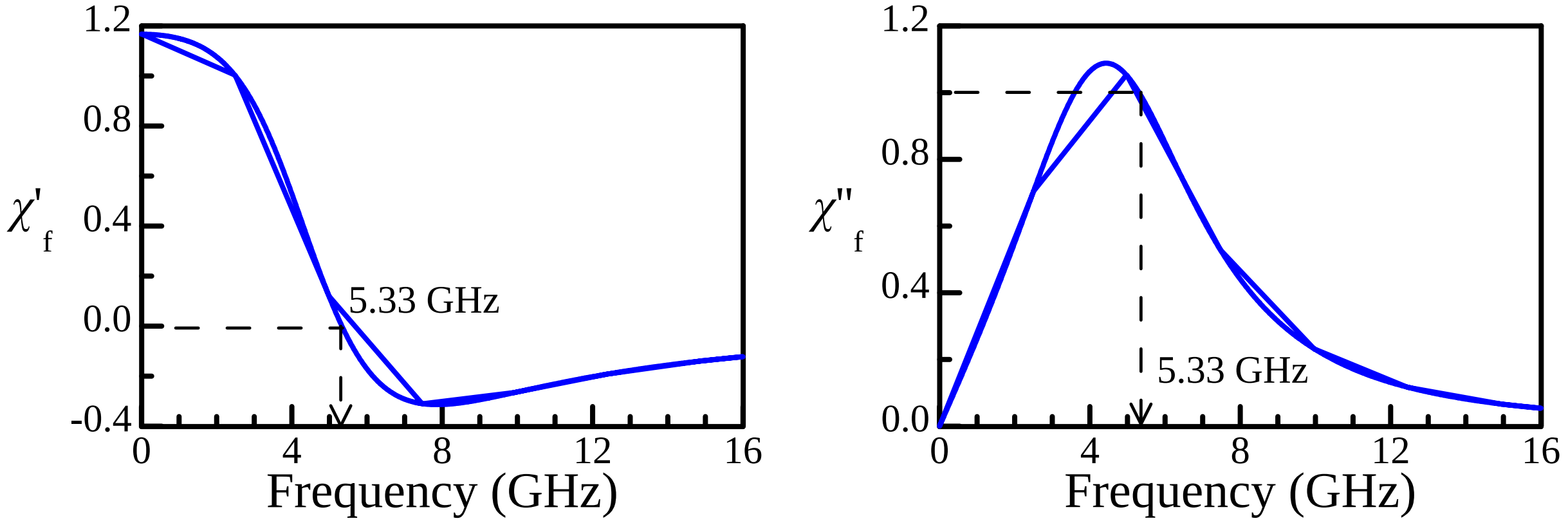

**Figure 4.** The complex magnetic susceptibility of the nano magnetite versus frequency, extracted by fitting Eq. (12).

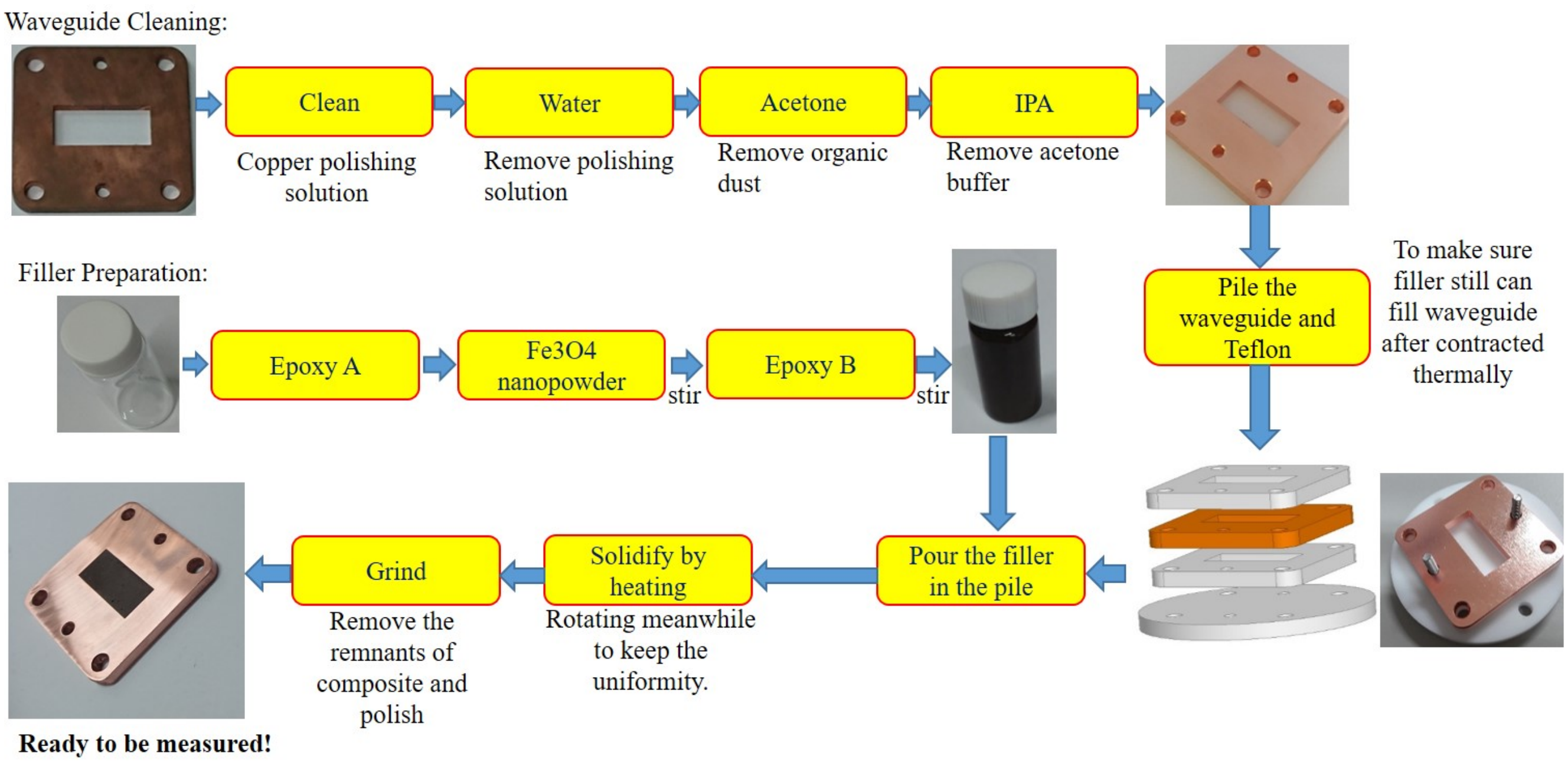

**Figure 5.** The sample preparing procedures: The waveguide is cleaned through four steps. The filler is prepared by mixing epoxy A, nano-magnetite powder, and epoxy B. Then, the mixture is poured into the mold consisting of the waveguide and other fixtures. The mold is dried for days, and the sample surface is ground and polished.

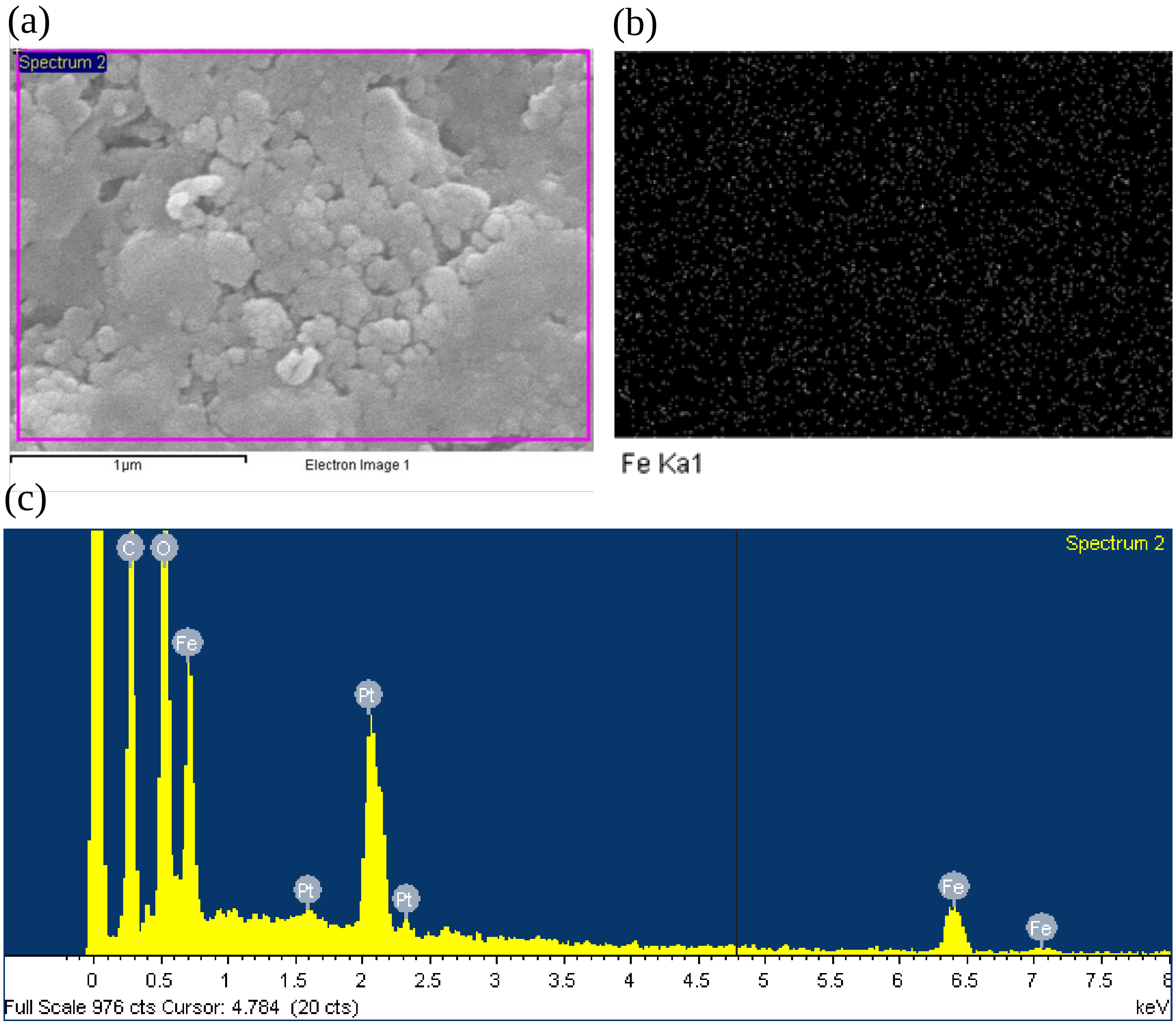

**Figure 6.** (a) The SEM image, (b) the corresponding distribution of iron through the Fe Ka X-ray microanalysis, and (c) the ingredient analysis using the X-ray emission spectrum.

| $\frac{2\mu_0 M_s^2 V}{3k_B T}$ | $\Gamma_f$ (GHz) | $\Theta_f$ (ns) |
|---|---|---|
| 1.167 | 4.55 | 0.16 |

**Table 1.** The corresponding parameters in Eq. (12) deduced by the fitting.

# Supplementary Information

**Che-Hao Chang[1], Shih-Chieh Su[2], Tsun-Hsu Chang[1, 2*], & Ching-Ray Chang[3**]**

[1]Interdisciplinary Program of Sciences, National Tsing Hua University, Hsinchu, Taiwan
[2]Department of Physics, National Tsing Hua University, Hsinchu, Taiwan
[3]Department of Physics, National Taiwan University, Taipei, Taiwan

### Part I. Differential form of the exponential memory kernel

In general, the correction of the master equation induced by the memory effect can be expressed as

$$\dot{\mathbf{m}}(t)=\int_{-\infty}^{t}K(t-\tau)(\overleftrightarrow{\mathbf{f}}_{\mathbf{0}}\mathbf{m}+qh\Gamma\mathbf{v})(\tau)d\tau, \tag{S1}$$

where $K$ is the memory kernel, representing how the history of the system influence the current system. Here, we set $K=\exp(-t/\Theta)/\Theta$ in the sense that the system has a finite memory time $\Theta$. Given this memory kernel, we can take the derivatives with respect to time on both sides of Eq. (S1). Then, we will get

$$\ddot{\mathbf{m}}(t)=K(0)(\overleftrightarrow{\mathbf{f}}_{\mathbf{0}}\mathbf{m}+qh\Gamma\mathbf{v})(t)+\int_{-\infty}^{t}\dot{K}(t-\tau)(\overleftrightarrow{\mathbf{f}}_{\mathbf{0}}\mathbf{m}+qh\Gamma\mathbf{v})(\tau)d\tau. \tag{S2}$$

Since $K(0)=1/\Theta$ and $\dot{K}(t-\tau)=-K(t-\tau)/\Theta$, Combining with Eq. (S1), we can furthermore rewrite Eq. (S2) as

$$\ddot{\mathbf{m}}(t)=\frac{1}{\Theta}((\overleftrightarrow{\mathbf{f}}_{\mathbf{0}}\mathbf{m}+qh\Gamma\mathbf{v})(t)-\dot{\mathbf{m}}(t)). \tag{S3}$$

and one can easily rewrite it as the form in Eq. (4).

### Part II. Interpretation of eigenvalues of $\overleftrightarrow{\mathbf{f}}_{\mathbf{0}}$

Note the diagonalization of $\overleftrightarrow{\mathbf{f}}_{\mathbf{0}}$ tells us that $m_1'=m_1-(m_2+m_3+m_4)$ is corresponding to the eigenvalue $-6\Gamma$, while $m_2'=m_1+3m_2-m_3-m_4$, $m_3'=m_1-m_2+3m_3-m_4$, and $m_4'=m_1-m_2-m_3+3m_4$ are corresponding to the eigenvalue $-2\Gamma$. The master equation of $m_n'$'s when $h=0$ and $\Theta=0$ is

$$\begin{bmatrix}\dot{m}_1'\\ \dot{m}_2'\\ \dot{m}_3'\\ \dot{m}_4'\end{bmatrix}=\begin{bmatrix}-6\Gamma & 0 & 0 & 0\\ 0 & -2\Gamma & 0 & 0\\ 0 & 0 & -2\Gamma & 0\\ 0 & 0 & 0 & -2\Gamma\end{bmatrix}\begin{bmatrix}m_1'\\ m_2'\\ m_3'\\ m_4'\end{bmatrix}. \tag{S4}$$

Therefore, while $m_1'$ has a relaxation time $1/(6\Gamma)$, $m_2'$, $m_3'$ and $m_4'$ has a relaxation time $1/(2\Gamma)$.

The difference can be understood by the following interpretation. If we understand $m_n$ as the reduced magnetization along the corresponding direction, then what $m_1'$ represents is the difference between the reduced magnetization along $[111]$ and the sum of the reduced magnetizations along three adjacent directions. On the other hand, $m_2'$ can be rewritten as $m_2' = (m_2 - (-m_1)) + (m_2 - m_3) + (m_2 - m_4)$, where $-m_1$ represents the reduced magnetizations along $[\bar{1}\bar{1}\bar{1}]$. Therefore what $m_2'$ represents is the sum of the difference between the reduced magnetization along $[\bar{1}11]$ and three directions adjacent to the opposite direction. $m_3'$ and $m_4'$ just replace $[\bar{1}11]$ by $[1\bar{1}1]$ and $[11\bar{1}]$. The $\overleftrightarrow{\mathbf{f}}_\mathbf{0}$ matrix allows only direct transitions between adjacent states. For non-adjacent states, the transitions are indirect, and therefore it takes more time for the non-adjacent states' transitions than for the adjacent transitions. That why $m_1'$ has a shorter relaxation time than the others, since the difference between adjacent states disappears faster than that between non-adjacent states.

With the corresponding relaxation time, it is natural that the frequency response of $m_1'$ will be the form $m_1' = C_1(\mathbf{H})qh\Gamma/(6\Gamma - \Theta\omega^2 + i\omega)$ and the frequency response of other $m_n'$ will be the form $m_n' = C_n(\mathbf{H})qh\Gamma/(2\Gamma - \Theta\omega^2 + i\omega)$. To see why we only see $2\Gamma$ dependence in the total $m_r$, note Eq. (8) can also be written as

$$m_r = \frac{1}{4}[\cos\varphi_1 \quad \cos\varphi_2 \quad \cos\varphi_3 \quad \cos\varphi_4] \begin{bmatrix} 1 & 1 & 1 & 1 \\ -1 & 1 & 0 & 0 \\ -1 & 0 & 1 & 0 \\ -1 & 0 & 0 & 1 \end{bmatrix} \begin{bmatrix} m_1' \\ m_2' \\ m_3' \\ m_4' \end{bmatrix}. \tag{S4}$$

Hence, $m_1'$ contributes to $m_r$ as the form $(\cos\varphi_1 - \cos\varphi_2 - \cos\varphi_3 - \cos\varphi_4)m_1'/4$. However, $\cos\varphi_n$'s are related to each other by $\cos\varphi_1 = \cos\varphi_2 + \cos\varphi_3 + \cos\varphi_4$. As a result, $m_1'$ can not contribute to $m_r$ in the sense that the value of $m_r$ won't change before and after the corresponding transition. For example, when the applied field is along $[111]$, $\cos\varphi_1 = 1$ and $\cos\varphi_n = 1/3$ for the rest. Hence, if $m_{1,0} \to m_{1,0} - \Delta m$ by the corresponding transition, $m_{n,0} \to m_{n,0} + \Delta m$ for the rest and $m_r = m_{1,0} + (m_{2,0} + m_{3,0} + m_{4,0})/3$ before and after the transition.

**Part III. Understanding the memory effect using the RLC current**

An RLC current satisfies

$$\frac{L}{R}\ddot{q} + \dot{q} + \frac{1}{RC}q = \frac{V}{R}, \tag{S5}$$

where $L$ is the inductance; $R$ is the resistor; $C$ is the capacitance; $V$ is the applied voltage, and $q$ is the charge on the capacitor. As the form looks like Eq. (4), we may explain Eq. (4) by each term of the RLC current.

For $-\overleftrightarrow{\mathbf{f}}_\mathbf{0}\mathbf{m}$, when the applied voltage is DC, then $\mathbf{m}$ will converge to $-\overleftrightarrow{\mathbf{f}}_\mathbf{0}\mathbf{m} = qh\Gamma\mathbf{v}$ just as $q$ will converge to $q = VC$. Therefore $-\overleftrightarrow{\mathbf{f}}_\mathbf{0}$ acts as the character of the capacitance. In addition, as the supplementary information shows, the eigenvalues of $-\overleftrightarrow{\mathbf{f}}_\mathbf{0}$ represent relaxation time, just like $RC$.

For $\Theta\ddot{\mathbf{m}}$, it acts like $L/R$. To see why the memory effect has such a meaning, note an inductor acts as the role that it keeps current from changing. One can check the form of the memory effect in Eq. (S1) again:

$$\dot{\mathbf{m}}(t) = \int_{-\infty}^{t} K(t-\tau)(\overleftrightarrow{\mathbf{f}}_\mathbf{0}\mathbf{m} + qh\Gamma\mathbf{v})(\tau)d\tau. \tag{S6}$$

This equation tells us what the memory kernel does is to memorize the previous value of $\dot{\mathbf{m}}$ and try to keep $\dot{\mathbf{m}}$ to be the same, just as the inductor tries to keep $\dot{q}$ to be the same.